\documentclass[epj]{svjour}
\usepackage{amsmath,amssymb,graphicx,hyperref}
\newcommand{\dd}{\mathrm{d}}
\renewcommand{\footnote}[1]{}

\begin{document}

\title{Spectral densities of Wishart-L\'evy free stable random matrices}
\subtitle{Analytical results and Monte Carlo validation}
\author{Mauro Politi\inst{1,2} \and Enrico Scalas\inst{3}
\and Daniel Fulger \inst{2,3,4} \and Guido Germano\inst{2}
} 


\institute{
Department of Physics, University of Milan, Via Giovanni Celoria 16,
20133 Milano, Italy
\and
Department of Chemistry and WZMW, Computer Simulation Group,
Philipps-University Marburg, 35032 Marburg, Germany
\and
Department of Advanced Sciences and Technology, Laboratory on Complex Systems,
Amedeo Avogadro University of East Piedmont, Via Vincenzo Bellini 25 G,
15100 Alessandria, Italy
\and
Institute for Scientific Interchange, Complex Systems Lagrange Lab,
Viale Settimio Severo 65, 10133 Torino, Italy
}

\date{Received: date / Revised version: date}

\abstract{
Random matrix theory is used to assess the significance of weak correlations
and is well established for Gaussian statistics. However, many complex systems,
with stock markets as a prominent example, exhibit statistics with power-law
tails, that can be modelled with L\'evy stable distributions. We review
comprehensively the derivation of an analytical expression for the spectra of
covariance matrices approximated by free L\'evy stable random variables and
validate it by Monte Carlo simulation.
} 

\PACS{ 
{89.65.Gh} {Economics; econophysics, financial markets, business and management}
\and
{02.50.Ng} {Distribution theory and Monte Carlo studies} \and
{02.70.Uu} {Applications of Monte Carlo methods}
} 

\maketitle

\section{Introduction}
\label{sec:Intoduction}

The classical ensembles of random matrices play an important role in the
modelling of physical systems, in time series analysis and in other fields.
The first notion of a matrix ensemble in statistics was given in the 1920s by
Wishart for the purpose of correlation analysis \cite{Wishart1928}. Physicists
began to be interested in random matrices in the 1950s, when Wigner presented a
model of nuclear energy levels as eigenvalues of symmetric random matrices
$\mathbf{W}$ whose elements are random numbers drawn from a Gaussian
distribution $N(0,\sigma^2)$ \cite{Wigner1955}, or actually from any symmetric
distribution with a finite second moment \cite{Wigner1958}, e.g.\ equiprobable
$\pm1$ random numbers. With increasing matrix size the eigenvalue spectrum
tends to the semicircle law:
\begin{equation}\label{eq:semicircle}
\rho_\mathbf{W}(\lambda) = \frac{1}{2\pi\sigma^2}\sqrt{4\sigma^2-\lambda^2}.
\end{equation}
Wigner's data were based on neutron and proton scattering. Other applications
of random matrix theory in physics include classical and quantum chaos,
disordered systems, many-body quantum systems, quantum dots, quantum
chromodynamics, quantum gravity, supersymmetric field theory, string theory,
etc. In 1998 Guhr et al.\ wrote a review on many of these with more than
800 references \cite{Guhr1998}. In 2003 the Journal of Physics A dedicated a
special issue to random matrix theory \cite{Forrester2003}. Random matrices are
used in other fields too, e.g.\ operations research, for diverse problems as
bandwith efficiency in wireless communication \cite{Foschini1998,Moustakas2005}
or optimal aircraft boarding \cite{Bachmat2006,Steffen2008}. In correlation
analysis the theory of random matrices can be used to assess whether weak
correlations are significant or just noise. The mathematical link between
correlation matrices of time series and random matrices is the Wishart matrix
ensemble, that, together with the Wigner ensemble, is one of the standard tools
in the theory of random matrices. Recent introductions to the latter including
numerical aspects can be found in Refs.~\cite{Mehta2004,Edelman2005}. Since the
1990s econophysicists have employed random matrix theory for the analysis of
correlation in financial time series \cite{Laloux1999,Plerou1999,Bouchaud2000,%
Bouchaud2004,Tola2008,Daly2008}, with portfolio theory \cite{Markowitz1952,%
Sharpe1964} as one of the motivations; a particular attention is given to the
largest eigenvalues of the covariance matrix and the associated eigenvectors,
that correspond to the whole market and its sectors. Recently, random matrix
theory was used also for a correlation analysis of macroeconomic time series
\cite{Ormerod2008}.

Consider $i = 1, \ldots, N$ stochastic time series $x_{ij}$ observed at
synchronous times $t_j,\ j = 0,\ldots,T$. The data can be arranged in a
$N \times T$ matrix $\mathbf{M}$ of increments $m_{ij} = x_{ij} - x_{i,j-1}$,
where each row corresponds to a time series and each column to a sampling time.
Assuming that the average of the increments is zero, the Pearson estimator for
the covariance of two time series $i$ and $j$ is
\begin{equation}\label{eq:covariance}
c_{ij} = \frac{1}{T} \sum_{k=1}^T m_{ik} m_{jk}.
\end{equation}
The covariances of all pairs can be collected in a $N \times N$ symmetric
matrix
\begin{equation}\label{eq:covmatr}
\mathbf{C} = \frac{1}{T} \mathbf{M M}^\mathsf{T}.
\end{equation}
The covariance matrix $\mathbf{C}$ is also called Wishart matrix as it was
studied by him. One is often interested in testing the hypothesis that there
are no significant correlations. This can be done comparing the eigenvalue
spectrum of an empirical correlation matrix with the spectrum of a reference
matrix built with synthetic uncorrelated time series. If the matrix rows are
random walks whose increments are independent and identically distributed (iid)
normal deviates with standard deviation $\sigma$, the spectrum describing the
above null hypothesis in the limit for $N,\ T \rightarrow \infty$ with $m =
N/T$ is given analytically by the Mar\v{c}enko-Pastur law \cite{Marchenko1967}:
\begin{gather}\label{eq:Marchenko-Pastur}
\rho_\mathbf{C}(\lambda)
= \frac{\sqrt{(\lambda_+-\lambda)(\lambda-\lambda_-)}}{2\pi\sigma^2m\lambda}
\\ \lambda_\pm = \sigma^2\left(1 \pm \sqrt{m}\right)^2. \nonumber
\end{gather}
This result has been rediscovered a few times \cite{Edelman2005,Edelman1988,%
Bai1999}. Indeed, for a sufficiently large matrix the exact distribution of its
elements becomes less and less relevant, and the Mar\v{c}enko-Pastur law can be
obtained for iid increments drawn from any symmetric distribution with a finite
second moment $\sigma^2$. This effect was evident also in Wigner's studies of
matrices whose elements are binary random variables assuming the values $\pm 1$
with equal probability. In both the Wigner and Wishart ensembles the spectra of
large matrices converge to that of an infinite matrix (respectively the
semicircle law and the Mar\v{c}enko-Pastur law) as a consequence of a
generalised central limit theorem.

A practical use of Eq.~(\ref{eq:Marchenko-Pastur}) is that if the empirical
spectrum of data shows significant differences from the theoretical curve, then
it may be justified to reject the null hypothesis of no true correlations.
The details of the latter are then a separate issue. In principle it is
possibile to test not only correlation, but also any kind of suitable
assumption leading to a given shape of the expected spectrum, both
theoretically and numerically. Depending on the specific case one chooses a
suitable null hypothesis. For example, if the considered time series are the
log-prices of traded stocks, in a first approximation it is reasonable to test
the absence of true correlation with normally distributed log-returns
\cite{Laloux1999,Plerou1999,Clauset2007}. Another powerful approach requiring
less knowledge of the distribution of the increments is a bootstrap scheme that
consists in resampling the covariance matrix after random permutations of the
empirical time series. Since the reshuffling of the rows of $\mathbf{M}$
destroys any possible correlation, an absence of correlation among the original
time series requires that the eigenvalue spectrum of $\mathbf{C}$ does not
change. 

So far, the result given by Eq.~(\ref{eq:Marchenko-Pastur}) lies within
classical random matrix theory and requires iid matrix elements with finite
moments. In this work we are concerned with the Wishart-L\'evy ensemble as a
natural extension of the Wishart-Gaussian ensemble treated by the
Mar\v{c}enko-Pastur theory. The situation becomes more complicated if the
elements of $\mathbf{M}$ are distributed with power-law tails, as happens in
numerous physical, biological and economic data \cite{Clauset2007}. Stock
markets as well as many other complex systems exhibit a dynamics that results
in power-law tailed statistics. The Mar\v{c}enko-Pastur theory is not valid any
more when the second moment is not finite, and the corresponding spectral
densities cannot be obtained from a simple extension of Gaussian random matrix
theory. As a consequence of the central limit theorem for scale-free processes
the distribution of many of the above phenomena is usually assumed to be a
symmetric L\'evy $\alpha$-stable distribution, whose pdf is given most suitably
as the inverse Fourier (cosine) transform of its characteristic function:
\begin{eqnarray}\label{eq:levy}
L_\alpha (x) &=& \mathcal{F}^{-1}_k\left[e^{-|\gamma k|^\alpha}\right](x) \\
&=& \frac{1}{\pi} \int_0^\infty e^{-(\gamma k)^\alpha} \cos(x k)\, \dd k.
\nonumber
\end{eqnarray}
The second and higher moments of $L_\alpha (x)$ diverge for {$\alpha < 2$},
and for $\alpha \le 1$ even the first moment does not exist. If $\alpha = 2$
Eq.~(\ref{eq:levy}) gives a Gaussian with standard deviation $\sigma =
\sqrt{2}\gamma$. However, we shall see that the functional representation of
this distribution is not required in the derivation of the spectrum.

A matrix whose elements are iid samples from a stable density is called a
L\'evy matrix. A symmetric L\'evy matrix is called a Wigner-L\'evy matrix.
A symmetric matrix $\mathbf{C}$ built from a L\'evy matrix $\mathbf{M}$
according to the equation
\begin{equation}
\mathbf{C} = \frac{1}{T^{2/\alpha}} \mathbf{M M}^\mathsf{T}
\end{equation}
is called a Wishart-L\'evy matrix. Notice that the normalisation factor has
been generalised with respect to Eq.~(\ref{eq:covmatr}) to take into account
L\'evy $\alpha$-stable statistics. Sampling the elements from the probability
density function
\begin{equation}
f_X(x) = N^{2/\alpha} L_\alpha (N^{2/\alpha}x),
\end{equation}
the limiting spectrum becomes independent of the matrix size $N$
\cite{Cizeau1994}. It turns out that the spectra of these matrices have no
longer a finite support as in the semicircle and Mar\v{c}enko-Pastur laws and
are dominated by the behaviour of the power-law tail of $L_\alpha(x)$.

It was proposed to use the theory of free probability with its convenient
machinery leading to analytic results that could be obtained otherwise only by
means of a painful use of combinatorics. A free L\'evy stable random matrix has
a spectrum belonging to the class of free stable laws. The contemporary
physical and mathematical literature on correlation matrix analysis with
power-law tailed uncorrelated noise is very active also in the context of free
probability. Limiting the list to physics journals, the reader can consult
Refs.~\cite{Burda2001,Burda2002,Burda2003,Burda2004a,Burda2004b,Burda2004c,%
Burda2005,Burda2006a,Burda2006b,Burda2007,Bouchaud2007a,Bouchaud2007b,%
Biroli2007,Vivo2007}. For a review of free probability theory see
Ref.~\cite{Nica2006}. The Mar\v{c}enko-Pastur spectrum can be obtained as a
special case of this more general theory.

Our aim in this paper is to review comprehensively the analytic derivation of
the spectral density of free stable Wishart-L\'evy random matrices already
solved by Burda et al.\ \cite{Burda2001,Burda2002,Burda2004a,Burda2004b,%
Burda2004c,Burda2005,Burda2006a,Burda2006b,Burda2007} and, as a further step,
to validate numerically the analytic result by Monte Carlo simulation.
The rest of this paper is organised as follows. Sec.~\ref{sec:math} introduces
the mathematical background of free probability theory, whose objects are
elements of an algebra, usually an operator algebra, and may enjoy the property
of freeness. Sec.~\ref{sec:approximation} explains free stability and presents
an approximation for the Wishart-L\`evy covariance matrix of time series using
free stable random variables. An explanation of free stability is provided too.
Sec.~\ref{sec:analytical} derives in detail a transcendental equation, due to
Burda et al., whose solution gives the spectral density for the approximated
covariance matrix. Sec.~\ref{sec:MC} shows numerically the validity of this
equation comparing analytical and Monte Carlo results. A summary and an
appendix with computer code conclude the paper.

\section{Mathematical background}\label{sec:math}

A symmetric $N \times N$ matrix $\mathbf{X}$ has real eigenvalues $\lambda_1,
\ldots, \lambda_N$. The spectral density of $\mathbf{X}$ can be written as
\begin{equation}
\rho_\mathbf{X}(\lambda) = \frac{1}{N}\sum_{i=1}^N \delta(\lambda-\lambda_i),
\end{equation}
where it is assumed that the weight of each eigenvalue is the same and each
eigenvalue is counted as many times as its multiplicity. The resolvent matrix
\cite{Mikusinski2005} is defined as
\begin{equation}\label{eq:resolvent}
\mathbf{G}_\mathbf{X}(z) = (z\mathbf{1}-\mathbf{X})^{-1},\quad z \in\mathbb{C},
\end{equation}
where $\mathbf{1}$ is the $N \times N$ identity matrix. The Green function is
defined as
\begin{equation}\label{eq:greenf}
G_\mathbf{X}(z) = \frac{1}{N}\, \mathrm{tr}\, \mathbf{G}_\mathbf{X}(z),
\end{equation}
where the trace $\mathrm{tr}$ of a square matrix is defined as the sum of its
diagonal elements. If $\mathbf{X}$ is a random matrix, the above definition is
generalised including an expectation operator $\mathbb{E}$:
\begin{equation}\label{eq:greenfrandommatrix}
G_\mathbf{X}(z)= \frac{1}{N}\, \mathbb{E}[\mathrm{tr}\, \mathbf{G_X}(z)].
\end{equation}
The Green function contains the same information as the eigenvalues and the
eigenvalue density of $\mathbf{X}$ \cite{Bouchaud2000}. The Green function can
be written in terms of the eigenvalues of $\mathbf{X}$:
\begin{equation}
G_\mathbf{X}(z) = \frac{1}{N}\, \sum_{i=1}^N \frac{1}{z-\lambda_i}.
\end{equation}
This is a special case of the definition through the Cauchy transform of a
generic spectral density:
\begin{equation}\label{eq:cauchytr}
G_\mathbf{X}(z) = \int_{-\infty}^{+\infty}
\frac{1}{z-\lambda}\rho_\mathbf{X}(\lambda)\, \dd\lambda.
\end{equation}
By using the following representation of Dirac's $\delta$-function,
\begin{equation}
\frac{1}{x\pm i\epsilon}=\mathrm{PV}\left(\frac{1}{x}\right) \mp i\pi\delta(x),
\end{equation}
where $\mathrm{PV}$ denotes the principal value, the spectral density can be
obtained from the Green function:
\begin{equation}\label{eq:rhofromG}
\rho_\mathbf{X}(\lambda)= \lim_{\epsilon\to 0^+}\frac{1}{\pi} \mathrm{Im} [
G_\mathbf{X}(\lambda-i\epsilon)].
\end{equation}
This means that the eigenvalues follow from the discontinuities of
$G_\mathbf{X}(z)$ on the real axis.

Non-commutativity of matrices and, in general, of operators makes it difficult
to extend standard probability theory to matrix as well as operators spaces.
Among possible extensions of probability theory to operator spaces the
so-called free probability theory has the advantage that many results can be
deduced from well-known theorems on analytic functions \cite{Burda2006b}.

In order to explain the framework of free probability, let us start from
conventional classical probability. A probability space $(\Omega,\mathcal{F},
\mathrm{P})$ is a measure space, where $\Omega$ is the sample space,
$\mathcal{F}$ is a $\sigma$-algebra on $\Omega$, and $\mathrm{P} : \mathcal{F}
\to [0,1] \in \mathbb{R}$ is a non-negative measure on sets in $\mathcal{F}$
obeying Kolmogorov's axioms; $\omega \in \Omega$ is called an elemetary event,
$A \in \mathcal{F}$ is called an event. A random variable $X : \Omega \to
\mathbb{R}$ is a measurable function that maps elements from the sample space
to the real numbers, and thus elements from $\mathcal{F}$ to a Borel
$\sigma$-algebra $\Sigma$ on $\mathbb{R}$. The probability distribution of $X$
with respect to $\mathrm{P}$ is described by a measure $\mu_X$ on $(\mathbb{R},
\Sigma)$ defined as the image measure of $\mathrm{P}$: $\mu_X(B) = \mathrm{P}
[X^{-1}(B)]$, where $B$ is any Borel set and $X^{-1}(B) \subset \mathcal{F}$ is
the counter-image of $B$. The cumulative distribution function of $X$ is
$F_X(x) = \mu_X(X\le x)$. The expectation value for any bounded Borel function
$g : \mathbb{R} \to \mathbb{R}$ is
\begin{equation}
\mathbb{E}[g(X)] = \int_\mathbb{R} g(x) \mu_X(\dd x)
= \int_\mathbb{R} g(x)\, \dd F_X(x).
\end{equation}
If $F_X(s)$ is differentiable, the probability density function (pdf) of $X$ is
$f_X(x) = \dd F_X(x)/\dd x$.

This construction can be extended to non-commutative variables, e.g.\ matrices
or more in general operators. Let $\mathcal{A}$ denote a unital algebra over a
field $\mathbb{F}$, i.e.\ a vector space equipped with a bilinear product
$\circ: \mathcal{A}\times\mathcal{A} \to \mathcal{A}$ that has an identity
element $\mathbf{I}$. A tracial state on $\mathcal{A}$ is a positive linear
function $\tau: \mathcal{A} \to \mathbb{F}$ with the properties
$\tau(\mathbf{I}) = 1$ and $\tau(\mathbf{XY}) = \tau(\mathbf{YX})$ for every
$\mathbf{X,Y} \in \mathcal{A}$. The couple $(\mathcal{A},\tau)$ is called a
non-commutative probability space.

For our purposes $\mathcal{A} = \mathcal{B}(\mathcal{H})$, where
$\mathcal{B}(\mathcal{H})$ denotes the Banach algebra of linear operators on a
real separable Hilbert space $\mathcal{H}$. This is a $*$-algebra, as it is
equipped with an involution (the adjoint operation) $\mathbf{X} \mapsto
\mathbf{X}^* : \mathcal{B}(\mathcal{H}) \to \mathcal{B}(\mathcal{H})$.
%
%
Considering a self-adjoint operator $\mathbf{X} \in \mathcal{B}(\mathcal{H})$,
it is possible to associate a (spectral) distribution to $\mathbf{X}$ as in
classical probability. Thanks to the  Riesz representation theorem and the
Stone-Weierstrass theorem, there is a unique measure $\mu_\mathbf{X}$ on
$(\mathbb{R}, \Sigma)$ satisfying
\begin{equation}
\int_\mathbb{R} g(x) \mu_{\mathbf{X}}(\dd x) = \tau[g(\mathbf{X})]
\end{equation}
where $g:\mathbb{R}\to\mathbb{R}$ is any bounded Borel function
\cite{Nica2006}. Therefore we say that the distribution of $\mathbf{X}$
is described by the measure $\mu_{\mathbf{X}}$. For our purposes this measure
is equal to the spectral density $\rho_{\mathbf{X}}$ defined in
Eq.~(\ref{eq:rhofromG}). In random matrix theory the Wigner semicircle law has
the role of the Gaussian law in classical probability, and the
Mar\v{c}enko-Pastur law corresponds to the $\chi^2$ law.

Classically, independence between two random variables $X$ and $Y$ can be
defined requiring that for any couple of bounded Borel functions $f, g$
\begin{equation}
\mathbb{E}[(f(X)-\mathbb{E}[f(X)])(g(Y)-\mathbb{E}[g(Y)])] = 0.
\end{equation}
Analogously, two elements $\mathbf{X}$ and $\mathbf{Y}$ in a non-commutative
probability space are defined as free (of freely) independent with respect to
$\tau$, if for any couple of bounded Borel functions $f, g$
\begin{equation}
\tau[(f(\mathbf{X})-\tau[f(\mathbf{X})])(g(\mathbf{Y})-\tau[g(\mathbf{Y})])]=0.
\end{equation}
Defining freeness between more than two elements is a non-trivial
extension \cite{Barndorff-Nielsen2002}.

Generally, square $N \times N$ random matrices $\mathbf{X}$ are non-commutative
variables with respect to the function $\tau(\mathbf{X}) = (1/N)\, \mathbb{E}
[\mathrm{tr}\, \mathbf{X}]$, see Eq.~(\ref{eq:greenfrandommatrix}), but for
any given $N$ no pair of random matrices is free. Nevertheless two random
matrices $\mathbf{X,Y}$ can reach freeness asymptotically if for any integer
$n > 0$ and any set of non-negative integers $(\gamma_1,\ldots,\gamma_n)$ and
$(\beta_1,\ldots,\beta_n)$ for which in the limit $N \rightarrow \infty$
\begin{equation}
\tau(\mathbf{X}^{\gamma_1}) = \ldots = \tau(\mathbf{X}^{\gamma_n})
= \tau(\mathbf{Y}^{\beta_1}) = \ldots = \tau(\mathbf{Y}^{\beta_n}) = 0
\end{equation}
we have
\begin{equation}
\tau(\mathbf{X}^{\gamma_1}\,\mathbf{Y}^{\beta_1}\,\ldots\,\mathbf{X}^{\gamma_n}
\,\mathbf{Y}^{\beta_n}) = 0.
\end{equation}
This means that large random matrices can be good approximations of free
non-commutative variables.

Given an operator $\mathbf{X} \in \mathcal{B}(\mathcal{H})$, the following
functions are useful in deriving its spectral distribution $\mu_\mathbf{X}$:
\begin{enumerate}
\item \textit{Moment generating function,} defined as
\begin{equation}\label{eq:definw}
M_\mathbf{X}(z) = zG_\mathbf{X}(z) - 1.
\end{equation}
The name stems from the fact that, if the distribution of $\mathbf{X}$ has
finite moments of order $k$, $m_{\mathbf{X},k} =
\tau(\mathbf{X}^k)$,
\begin{equation}\label{eq:momentgen}
M_\mathbf{X}(z) = \sum_{k=1}^\infty \frac{m_{\mathbf{X},k}}{z^{k}}.
\end{equation}
This can be seen inserting the sum of the geometric series
\begin{equation}
\sum_{k=0}^{\infty} q^k = \frac{1}{1-q},\quad |q| < 1
\end{equation}
with $q = \lambda/|z|$ into Eq.~(\ref{eq:cauchytr}):
\begin{eqnarray}
G_\mathbf{X}(z) &=& \int_{-\infty}^{+\infty}
\frac{1}{z(1-\lambda/z)} \rho_\mathbf{X}(\lambda)\, \dd\lambda\\
&=& \int_{-\infty}^{+\infty} \frac{1}{z} \sum_{k=0}^\infty
\frac{\lambda^k}{z^k} \rho_\mathbf{X}(\lambda)\, \dd\lambda\\
&=& \sum_{k=0}^\infty \frac{1}{z^{k+1}} \int_{-\infty}^{+\infty}
\lambda^k \rho_\mathbf{X}(\lambda)\, \dd\lambda\\
&=& \sum_{k=0}^\infty \frac{m_{\mathbf{X},k}}{z^{k+1}}.
\end{eqnarray}

\item \textit{$R$-transform.} In classical probability the pdf of the sum of
two independent random variables $X+Y$ is equal to the convolution of the
individual pdfs, i.e.\
\begin{equation}\label{eq:convolution}
f_{X+Y}(x) = (f_X*f_Y)(x).
\end{equation}
The convolution is done conveniently in Fourier space, where it becomes a
multiplication: the characteristic function
\begin{equation}
\hat{f}_{X+Y}(k) = \int_\mathbb{R} f_{X+Y}(x) e^{ikx}\, \dd x
\end{equation}
of $X+Y$ is the product of the characteristic functions of $X$ and $Y$,
\begin{equation}
\hat{f}_{X+Y}(k) = \hat{f}_X(k)\hat{f}_Y(k),
\end{equation}
and the cumulant generating function of $X+Y$ is the sum of the cumulant
generating functions of $X$ and $Y$:
\begin{equation}
\log\hat{f}_{X+Y}(k) = \log\hat{f}_X(k) + \log\hat{f}_Y(k).
\end{equation}
The free analogue of the cumulant generating function is the $R$-transform
invented by Voiculescu \cite{Nica2006,Voiculescu1986,Bercovici1993} as part of
the functional inverse of the Green function:
\begin{equation}\label{Rtransfrom}
G_{\mathbf{X}} \left( R_{\mathbf{X}} (z) + \frac{1}{z} \right) = z.
\end{equation}
The $R$-transform for the sum of two free operators is the sum of their
$R$-transforms:
\begin{equation}\label{Rtransfromaditive}
R_{\mathbf{X} + \mathbf{Y}}(z) = R_{\mathbf{X}}(z) + R_{\mathbf{Y}}(z).
\end{equation}
The free analogue of convolution is indicated with the symbol $\boxplus$:
\begin{equation}
\mu_\mathbf{X+Y} = \mu_\mathbf{X} \boxplus \mu_\mathbf{Y}.
\end{equation}
This is computed through $R_\mathbf{X}$, given the connection between the Green
function $G_\mathbf{X}$ and the spectral distribution $\mu_\mathbf{X}$. Other
definitions of the $R$-transform were proposed later.

\item \textit{Blue function.} It is convenient to introduce also an 
inverse of the Green function $G_{\mathbf{X}}(z)$, called Blue function as a
pun \cite{Janik1997}:
\begin{equation}\label{eq:BlueFunction}
G_{\mathbf{X}}(B_{\mathbf{X}}(z)) = B_{\mathbf{X}} (G_{\mathbf{X}}(z)) = z.
\end{equation}
The Blue function is related to the $R$-transform by
\begin{equation}\label{eq:R2B}
B_{\mathbf{X}}(z) = R_{\mathbf{X}}(z) + \frac{1}{z} .
\end{equation}

\item \textit{$S$-transform.} In the same fashion as the $R$-transform for the
sum, another transform allows to compute the spectral distribution of the
product of two operators from their individual spectral distributions:
\begin{equation}\label{eq:Stransform}
S_\mathbf{X}(z) = \frac{1+z}{z}\chi_\mathbf{X}(z),
\end{equation}
where $\chi_\mathbf{X}(z)$ is defined through
\begin{equation}\label{eq:chidef}
\chi_\mathbf{X}(zG_\mathbf{X}(z)-1) = \frac{1}{z}.
\end{equation}
For $\mathbf{X} \neq \mathbf{Y}$ the $S$-transform of the product is the
product of the individual $S$-transforms:
\begin{equation}\label{eq:S_XY}
S_\mathbf{XY}(z) = S_\mathbf{X}(z)S_\mathbf{Y}(z).
\end{equation}
As the $R$-transform allows to compute the free additive convolution $\boxplus$,
the $S$-transform leads to the free multiplicative convolution $\boxtimes$:
\begin{equation}
\mu_\mathbf{XY} = \mu_\mathbf{X} \boxtimes \mu_\mathbf{Y}.
\end{equation}

\end{enumerate}

\section{Free stable random variables and the Wishart-L\'evy ensemble}\label{sec:approximation}

Let $\mathbf{P}$ be the matrix projector of size $T \times T$, with $N$ ones
in arbitrary positions on the diagonal and all the other elements zero, e.g.:
\begin{equation}\label{eq:P}
\mathbf{P} = \mathrm{diag}(\dots,1,1,\dots,0,1,0,0,1,\dots,1,0,\dots).
\end{equation}
Let $\mathbf{\Lambda}$ be a (large) $T \times T$ matrix with a free stable
spectral distribution. This property is the analogue of classical stability.
The sum of two free non-commutative $\mu$-distributed variables results in a
new $\mu$-distributed variable. The Wishart matrix ensemble of size $N \times
N$ defined in Eq.~(\ref{eq:covmatr}) can be approximated using the $N \times T$
matrix $\mathbf{M}/T^{1/\alpha}$ obtained from $\mathbf{P \Lambda}$ if only the
$N$ non-zero rows are considered \cite{Burda2001,Burda2002,Burda2004a,%
Burda2004b,Burda2004c,Burda2005,Burda2006a,Burda2006b,Burda2007}. Indicating
this operation with curly braces, the approximation reads
\begin{equation}\label{eq:corrPLLP}
\mathbf{C} = \frac{1}{T^{2/\alpha}} \mathbf{M M}^\mathsf{T} \simeq 
\{\mathbf{P\Lambda}\} \{\mathbf{\Lambda}^\mathsf{T}\mathbf{P}\}.
\end{equation}
The former equation is justified by very good results, both analytic and
numeric, in a similar approach for Wigner-L\'evy matrices \cite{Burda2007}.

Once we know the domain of attraction for one specific classical stable
distribution, we can expect that a sum of iid random numbers, e.g.\
$Z = (1/\mathcal{N}_n) \sum_{i=1}^n Z_i$ with some suitable normalisation
$\mathcal{N}_n$, converges to their attractor for large $n$. If $Z_i$ are
independent elements of random matrices, as in Ref.~\cite{Laloux1999}, each of
them tends to a stable law under matrix addition. However, for free stability
we must consider random matrices as a whole, and a different procedure is
needed. A fundamental point is a property discussed by Bercovici 
and Pata \cite{Bercovici1999}, that can be summarized as follows. 
If $\mathcal{D}_\mathrm{c}(\mu_\mathrm{c})$ and
$\mathcal{D}_\mathrm{f}(\mu_\mathrm{f})$ are the domains of attraction of the
stable laws $\mu_\mathrm{c}$ and $\mu_\mathrm{f}$ in classical and free
probability respectively, a distribution
$\nu \in \mathcal{D}_\mathrm{c}(\mu_\mathrm{c}) \Leftrightarrow
\nu \in \mathcal{D}_\mathrm{f}(\mu_\mathrm{f})$. In other words, if we are able
to recognise the classical attractor $\mathcal{D}_\mathrm{c}$ of a distribution
$\nu$, we also know its free attractor $\mathcal{D}_\mathrm{f}$. Moreover, one
and only one free stable distribution corresponds to any set of parameter
values characterising a classically stable distribution. The spectrum of a
Wigner-L\'evy matrix is symmetric with the same tail index $\alpha$ of its
entries, i.e.\ it belongs to the domain of attraction of a well-recognised
classical stable law. This means that the sum of sufficiently many free
non-commutative variables with this spectrum converges to a non-commutative
variable with a stable distribution.

Another property discussed in Refs.~\cite{Nica2006,Speicher1993,Pastur2000} can
be summarised for our purpose as follows. Considering two $N \times N$ matrices
$\mathbf{L}_i$ and $\mathbf{L}_j$ with $i \neq j$ and two independent random
orthogonal $N \times N$ matrices $\mathbf{O}_i$ and $\mathbf{O}_j$, the
matrices $\mathbf{O}_i \mathbf{L}_i\mathbf{O}_i^\mathsf{T}$ and $\mathbf{O}_j
\mathbf{L}_j \mathbf{O}_j^\mathsf{T}$ are free in the limit $N \rightarrow
\infty$. These properties together with the observation that $\mathbf{L}_i$ and
$\mathbf{O}_i \mathbf{L}_i \mathbf{O}_i^\mathsf{T} $ have the same spectrum
justify the equation~\cite{Burda2007}
\begin{equation}\label{eq:sum}
\mathbf{\Lambda} \simeq \frac{1}{(TR)^{1/\alpha}}
\sum_{i=1}^R \mathbf{O}_i \mathbf{L}_i \mathbf{O}_i^\mathsf{T}.
\end{equation}
This means that a free stable non-commutative variable can be approximated
adding randomly rotated classical L\'evy random matrices.

To generate L\'evy matrices we use the Chambers-Mal\-lows-Stuck algorithm
\cite{Chambers1976,McCulloch1996}: a random number $X$ drawn from the symmetric
L\'evy $\alpha$-stable pdf, Eq.~(\ref{eq:levy}), can be obtained from two
independent uniform random numbers $U, V \in (0,1)$ through the transformation
\begin{equation}\label{eq:chambers}
X = \gamma \left(\frac{-\log U \cos\Phi}{\cos((1-\alpha)\Phi)}
\right)^{1-\frac{1}{\alpha}} \frac{\sin(\alpha\Phi)}{\cos\Phi},
\end{equation}
where $\Phi = \pi(V-1/2)$. For $\alpha = 2$ Eq.~(\ref{eq:chambers}) reduces to
$X = 2\gamma\sqrt{-\log U}\sin\Phi$, i.e.\ the Box-Muller method for
Gaussian deviates with standard deviation $\sigma = \sqrt{2}\gamma$.

The QR-decomposition of a $T \times T$ matrix $\mathbf{H}$ with random Gaussian
entries yields
\begin{equation}
\mathbf{H} = \mathbf{O}\,\mathbf{U},
\end{equation}
where $\mathbf{O}$ is random orthogonal and $\mathbf{U}$ is upper (or right)
triangular. For alternative methods to obtain a random orthogonal matrix see
Ref.~\cite{Diaconis1987} and references therein.

\section{The analytical spectrum}\label{sec:analytical}

The moment generating function of the $T \times T$ matrix $\mathbf{D} =
\mathbf{\Lambda} \mathbf{P} \mathbf{\Lambda}^\mathsf{T}$ satisfies the
transcendental equation \cite{Burda2001,Burda2002,Burda2004a,Burda2006b}
\begin{equation}\label{eq:Burda}
-\exp\left(i\frac{2\pi}{\alpha} \right)\,z\,M_\mathbf{D}^{2/\alpha}(z)
= (M_\mathbf{D}(z)+1)(M_\mathbf{D}(z)+m),
\end{equation}
which can be solved analytically for a few special values of $\alpha = 1/4,\
1/3,\ 1/2,\ 2/3,\ 3/4,\ 1,\ 4/3,\ 3/2,\ 2$; the solution was published for
$\alpha = 1$ \cite{Burda2002}. The equation can be solved numerically for other
values, see the Appendix. Actually, we are interested in the spectrum of the
approximation of $\mathbf{C}$ provided by the rhs of Eq.~(\ref{eq:corrPLLP}),
but the Green functions of the matrices $\mathbf{D}$ and $\mathbf{C}$ are
related by the equation \cite{Burda2006b}
\begin{equation}
G_\mathbf{D}(z) = m^2\, G_\mathbf{C}(mz) + \frac{1-m}{z},
\end{equation}
whence, noticing that $m\,G_\mathbf{C}(mz) = G_\mathbf{C}(z)$,
\begin{equation}\label{eq:M_DtoM_C}
M_\mathbf{D}(z) = z\, G_\mathbf{D}(z) - 1 = m\,z\,G_\mathbf{C}(z) - m
= m \,M_\mathbf{C}(z).
\end{equation}
In the following we will explain in detail the route that leads to
Eq.~(\ref{eq:Burda}) and then to the desired spectral density $\rho_\mathbf{C}
(\lambda)$.

As in classical probability stable laws have an analytic form for their Fourier
transform, free stable laws have an analytic form for their Blue transform
\cite{Burda2007,Barndorff-Nielsen2002,Bercovici1999,Voiculescu1991}:
\begin{equation}\label{eq:stable}
B_\mathbf{\Lambda}(z;\alpha)= a + b z^{\alpha-1} + \frac{1}{z}.
\end{equation}
The parameter $a$ accounts for a horizontal shift in the distribution of the
matrix elements and can be set to zero without loss of generality. The
parameter $b$ depends on the distribution; for the symmetric L\'evy
$\alpha$-stable pdf, Eq.~(\ref{eq:levy}), it has the value \cite{Burda2004a}
\begin{equation}\label{eq:bfactor}
b = e^{i\pi(\alpha/2-1)}.
\end{equation}
As discussed in the previous section, given an index $\alpha \in (0,2]$,
$B_\mathbf{\Lambda} (z;\alpha)$ indirectly but precisely defines the attractor
law for the sum of free variables with $\alpha$-tailed spectral distribution.
Since free probability theory is exact only in the large size limit $T, N
\rightarrow \infty,\ N/T = m$, the only variables that define the model are
$\alpha$ and $m$.

Rewriting Eq.~(\ref{eq:stable}) with $G_\mathbf{\Lambda}(z)$ in place of $z$
and using Eq.~(\ref{eq:BlueFunction}) yields
\begin{equation}
b \, G_\mathbf{\Lambda}^{\alpha-1}(z) + G_\mathbf{\Lambda}^{-1}(z) = z,
\end{equation}
which is equivalent to
\begin{equation}\label{eq:geq}
b \, G_\mathbf{\Lambda}^\alpha(z) + z G_\mathbf{\Lambda}(z) + 1 = 0,
\quad G_\mathbf{\Lambda}(z) \neq 0.
\end{equation}

In Sec.~\ref{sec:math} we established calculation rules with the help of which
the solution of our specific problem can be put together piece by piece.
First notice that thanks to Eq.~(\ref{eq:S_XY}), if for simplicity from now on
we substitute $\mathbf{\Lambda}$ with its symmetrised counterpart
$(\mathbf{\Lambda + \Lambda}^\mathsf{T})/2$ so that $\mathbf{\Lambda =
\Lambda}^\mathsf{T}$,
\begin{equation}
S_{\mathbf{\Lambda P \Lambda}} = S_{\mathbf{\Lambda}} S_{\mathbf{P \Lambda}}
= S_{\mathbf{\Lambda}} S_{\mathbf{\Lambda P}} = S_{\mathbf{\Lambda \Lambda P}}
= S_{\mathbf{\Lambda}^2 \mathbf{P}}.
\end{equation}
For the $S$-transform of the matrix product $\mathbf{\Lambda}^2$ we also
require the Green function. The desired relation is a consequence of the fact
that the spectral measure for free L\'evy $\alpha$-stable operators in the
Wigner ensemble is symmetric \cite{Janik1997}:
\begin{gather}
\rho_\mathbf{\Lambda}(\lambda) = \rho_\mathbf{\Lambda}(-\lambda) \\
G_{\mathbf{\Lambda}}(z) = G_{-\mathbf{\Lambda}}(z).
\end{gather}
The Green function of $\mathbf{\Lambda}^2$ can be expressed in terms of the
Green function of $\mathbf{\Lambda}$ exploiting the Cauchy transform 
representation 
and the previous symmetry:
\begin{eqnarray}
G_{\mathbf{\Lambda}^2}(z) & = & \int_{-\infty}^{+\infty}\frac{1}{z-\lambda^2}
\rho_\mathbf{\Lambda}(\lambda)\, \dd\lambda \nonumber \\
& = & \int_{-\infty}^{+\infty} \left[ \frac{1}{2\sqrt{z}}
\left( \frac{1}{\sqrt{z}-\lambda} +  \frac{1}{\sqrt{z}+\lambda} \right)
\right] \rho_\mathbf{\Lambda}(\lambda)\, \dd\lambda \nonumber \\
& = & \frac{1}{2\sqrt{z}} \left( G_{\mathbf{\Lambda}}(\sqrt{z})
+ G_{-\mathbf{\Lambda}}(\sqrt{z}) \right) \nonumber \\ \label{eq:gsquare}
& = & \frac{1}{\sqrt{z}} G_{\mathbf{\Lambda}}(\sqrt{z}).
\end{eqnarray}

The next piece in the composition of the solution is the $S$-transform of the
projector $\mathbf{P}$, which requires its Green function too. Inserting
the spectral density of $\mathbf{P}$,
\begin{equation}
\rho_\mathbf{P}(\lambda) = m\delta(\lambda-1) + (1-m)\delta(\lambda),
\end{equation}
into the definition of the Green function of $\mathbf{P}$ as a Cauchy transform
yields
\begin{eqnarray}
G_\mathbf{P}(z) & = & \int\frac{1}{z-\lambda}\rho_\mathbf{P}(\lambda)\,
\dd\lambda \nonumber \\ & = & \nonumber \int\frac{1}{z-\lambda}
[m\delta(\lambda-1) + (1-m)\delta(\lambda)]\,\dd\lambda \\
& = & \frac{m}{z-1}+\frac{1-m}{z}.
\end{eqnarray}
The moment generating function $M_\mathbf{P}(z) = z G_\mathbf{P}(z) - 1$ and
the definition of the $S$-transform finally give
\begin{equation}
S_\mathbf{P} (z) = \frac{z+1}{z+m}.
\end{equation}

Rewriting Eq.~(\ref{eq:geq}) with $\sqrt{z}$ in place of $z$,
\begin{equation}
b\, G_\mathbf{\Lambda}^\alpha(\sqrt{z})
- \sqrt{z} G_\mathbf{\Lambda}^2(\sqrt{z}) + 1 = 0,
\end{equation}
and inserting Eq.~(\ref{eq:gsquare}) yields
\begin{equation}\label{eq:eqnresolv1}
b\, z^{\alpha/2} G_{\mathbf{\Lambda}^2}^\alpha(z)
- z G_{\mathbf{\Lambda}^2}(z) + 1 = 0.
\end{equation}
Observing that from Eq.~(\ref{eq:chidef})
\begin{equation}
z = \frac{1}{\chi_{\mathbf{\Lambda}^2}(zG_{\mathbf{\Lambda}^2}(z)-1)}
\equiv \frac{1}{\chi_{\mathbf{\Lambda}^2}},
\end{equation}
Eq.~(\ref{eq:eqnresolv1}) becomes
\begin{equation}\label{eq:eqnresolv2}
b\, \chi^{-\alpha/2} G_{\mathbf{\Lambda}^2}^\alpha
\left(\frac{1}{\chi_{\mathbf{\Lambda}^2}}\right)
- \frac{1}{\chi_{\mathbf{\Lambda}^2}} G_{\mathbf{\Lambda}^2}
\left(\frac{1}{\chi_{\mathbf{\Lambda}^2}}\right) + 1 = 0.
\end{equation}
Because from Eq.~(\ref{eq:Stransform}) it follows that
\begin{equation}\label{eq:blatozeta}
\frac{1}{\chi_{\mathbf{\Lambda}^2}} G_{\mathbf{\Lambda}^2}
\left(\frac{1}{\chi_{\mathbf{\Lambda}^2}}\right) - 1 = z,
\end{equation}
Eq.~(\ref{eq:eqnresolv2}) can be simplified to
\begin{equation}
b \chi_{\mathbf{\Lambda}^2}^{-\alpha/2} G_{\mathbf{\Lambda}^2}^{\alpha}
\left(\frac{1}{\chi_{\mathbf{\Lambda}^2}}\right) = z.
\end{equation}
Multiplying both sides by $\chi_{\mathbf{\Lambda}^2}^{-\alpha/2}/b$ yields
\begin{equation}
\chi_{\mathbf{\Lambda}^2}^{-\alpha} G_{\mathbf{\Lambda}^2}^{\alpha}
\left(\frac{1}{\chi_{\mathbf{\Lambda}^2}}\right)
= \frac{z}{b} \chi_{\mathbf{\Lambda}^2}^{-\alpha/2};
\end{equation}
then subtracting and adding 1,
\begin{equation}
\left(\frac{1}{\chi_{\mathbf{\Lambda}^2}} G_{\mathbf{\Lambda}^2}
\left(\frac{1}{\chi_{\mathbf{\Lambda}^2}}\right) -1 +1 \right)^{\alpha}
= \frac{z}{b}\chi_{\mathbf{\Lambda}^2}^{-\alpha/2},
\end{equation}
and inserting again Eq.~(\ref{eq:blatozeta}) gives
\begin{equation}
(z+1)^\alpha = \frac{z}{b}\chi_{\mathbf{\Lambda}^2}^{-\alpha/2},
\end{equation}
which can be written as
\begin{equation}
\chi_{\mathbf{\Lambda}^2}
= \frac{1}{(z+1)^2} \left(\frac{z}{b}\right)^{2/\alpha}.
\end{equation}
Now, using the definition of the $S$-transform and the result
\begin{equation}
S_{\mathbf{\Lambda}^2} =
\frac{1+z}{z} \chi_{\mathbf{\Lambda}^2}
= \frac{1}{z(1+z)} \left(\frac{z}{b}\right)^{2/\alpha},
\end{equation}
which can be used to write $S_\mathbf{D}$, the $S$-transform of the Wishart
matrix on the rhs of Eq.~(\ref{eq:corrPLLP}) is
\begin{equation}
S_{\mathbf{P} \mathbf{\Lambda}^2} = S_{\mathbf{P}} S_{\mathbf{\Lambda}^2}
=\frac{1}{z(m+z)} \left(\frac{z}{b}\right)^{2/\alpha}\,.
\end{equation}

This result is the starting point for the way back. Re-applying the definition
of the $S$-transform we can write
\begin{equation}
\chi_{\mathbf{\Lambda}^2 \mathbf{P}} = \frac{z}{z+1} S_{\mathbf{\Lambda}^2
\mathbf{P}} = \frac{1}{(z+1)(z+m)} \left(\frac{z}{b}\right)^{2/\alpha}
\end{equation}
and
\begin{equation}\label{eq:wishartfinal}
\chi_{\mathbf{\Lambda}^2 \mathbf{P}}^{-1} = (z+1)(z+m)
\left(\frac{z}{b}\right)^{-2/\alpha}.
\end{equation}
Together with $M_\mathbf{D}(z)=z\, G_\mathbf{D}(z)-1$ this allows to substitute
$\chi_\mathbf{D}(M_\mathbf{D}(z))=1/z$ and $M_\mathbf{D}(1/\chi_\mathbf{D}(z))
= z$. Notice that we changed the index $\mathbf{\Lambda}^2 \mathbf{P}$ to
$\mathbf{D}$ to emphasise our goal. So we can finally write
\begin{equation}\label{eq:BurdaAlgebSolution}
z = (M_\mathbf{D}(z)+1)(M_\mathbf{D}(z)+m)
\left(\frac{M_\mathbf{D}(z)}{b}\right)^{-2/\alpha}.
\end{equation}
Inserting Eq.~(\ref{eq:M_DtoM_C}) yields the corresponding equation for
$\mathbf{C}$:
\begin{equation}
z = (m\,M_\mathbf{C}(z)+1)(m\,M_\mathbf{C}(z)+m)
\left( \frac{m\,M_\mathbf{C}(z)}{b} \right)^{-2/\alpha};
\end{equation}
gathering $m$:
\begin{equation}\label{eq:transcendentalfinal}
z = m^{2-2/\alpha} (M_\mathbf{C}(z)+1/m)(M_\mathbf{C}(z)+1)
\left(\frac{M_\mathbf{C}(z)}{b} \right)^{-2/\alpha}.
\end{equation}
From Eq.~(\ref{eq:definw}) and from the relation between the moment generating
function and the spectrum we finally obtain
\begin{equation}\label{eq:spectrumfinal}
\rho_\mathbf{C}(\lambda)
= \frac{1}{\pi\lambda}\mathrm{Im}[M_\mathbf{C}(\lambda+i 0^-)].
\end{equation}
Inserting $b$ from Eq.~(\ref{eq:bfactor}) and rearranging,
Eq.~(\ref{eq:BurdaAlgebSolution}) takes the form anticipated in
Eq.~(\ref{eq:Burda}). Returning to the motivation of the paper, the result
described by Eq.~(\ref{eq:transcendentalfinal}) must be considered an
approximation of the curve corresponding to the null hypothesis of absence of
correlation in time series with fat-tailed increments.

\section{Monte Carlo validation}\label{sec:MC}

\begin{figure*}[htbp]
\begin{center}
\includegraphics[width=0.49\textwidth]{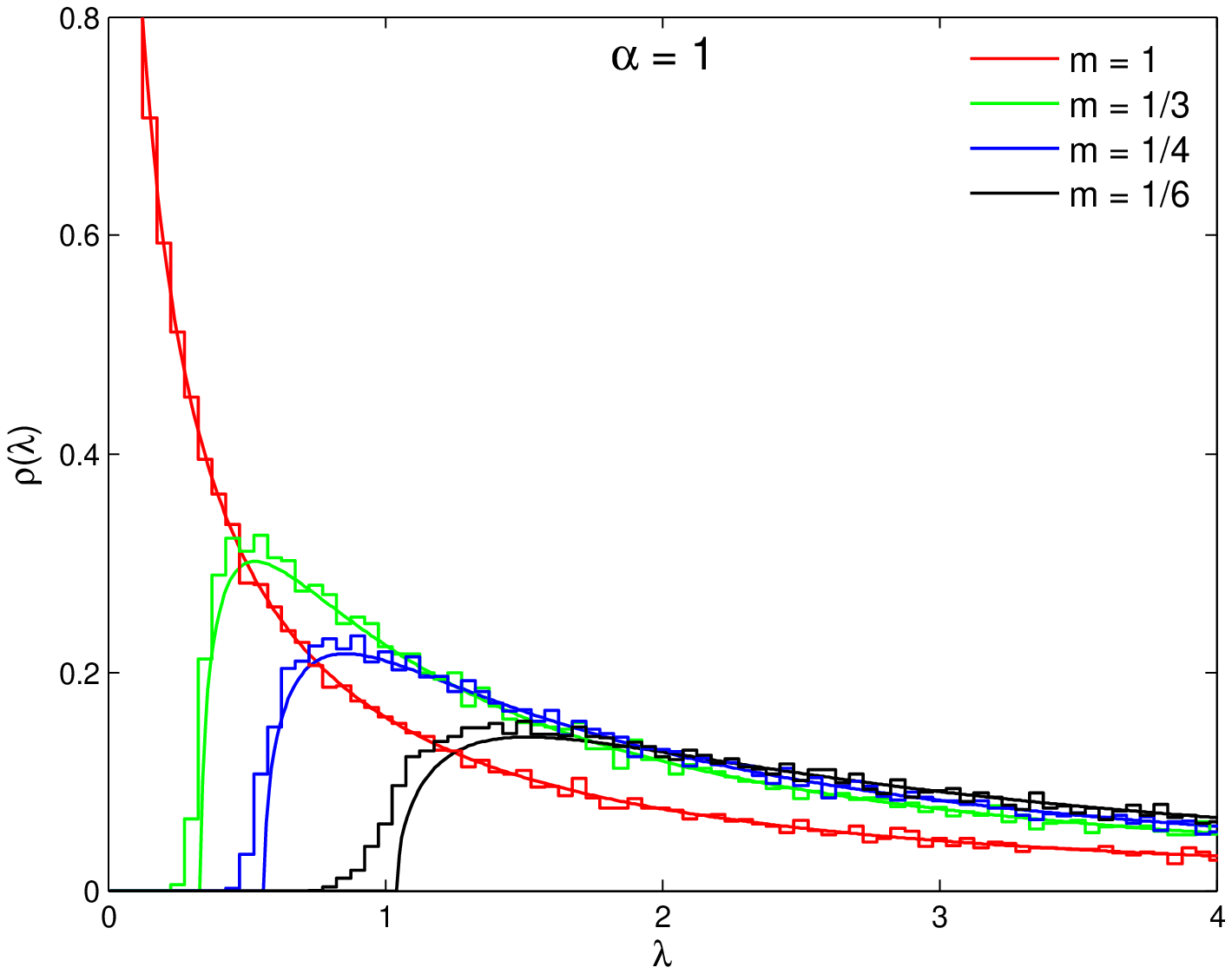}
\includegraphics[width=0.49\textwidth]{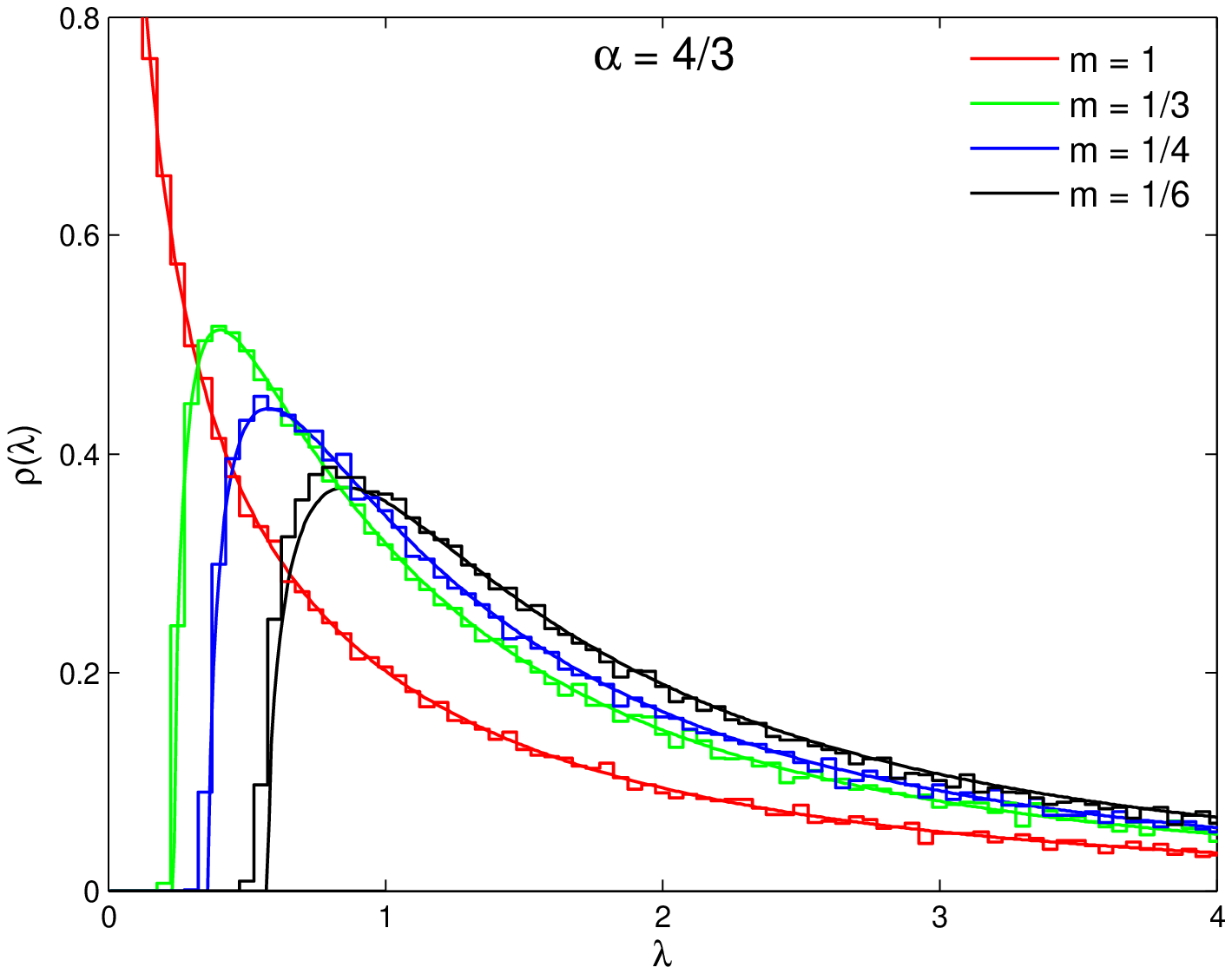}\\
\includegraphics[width=0.49\textwidth]{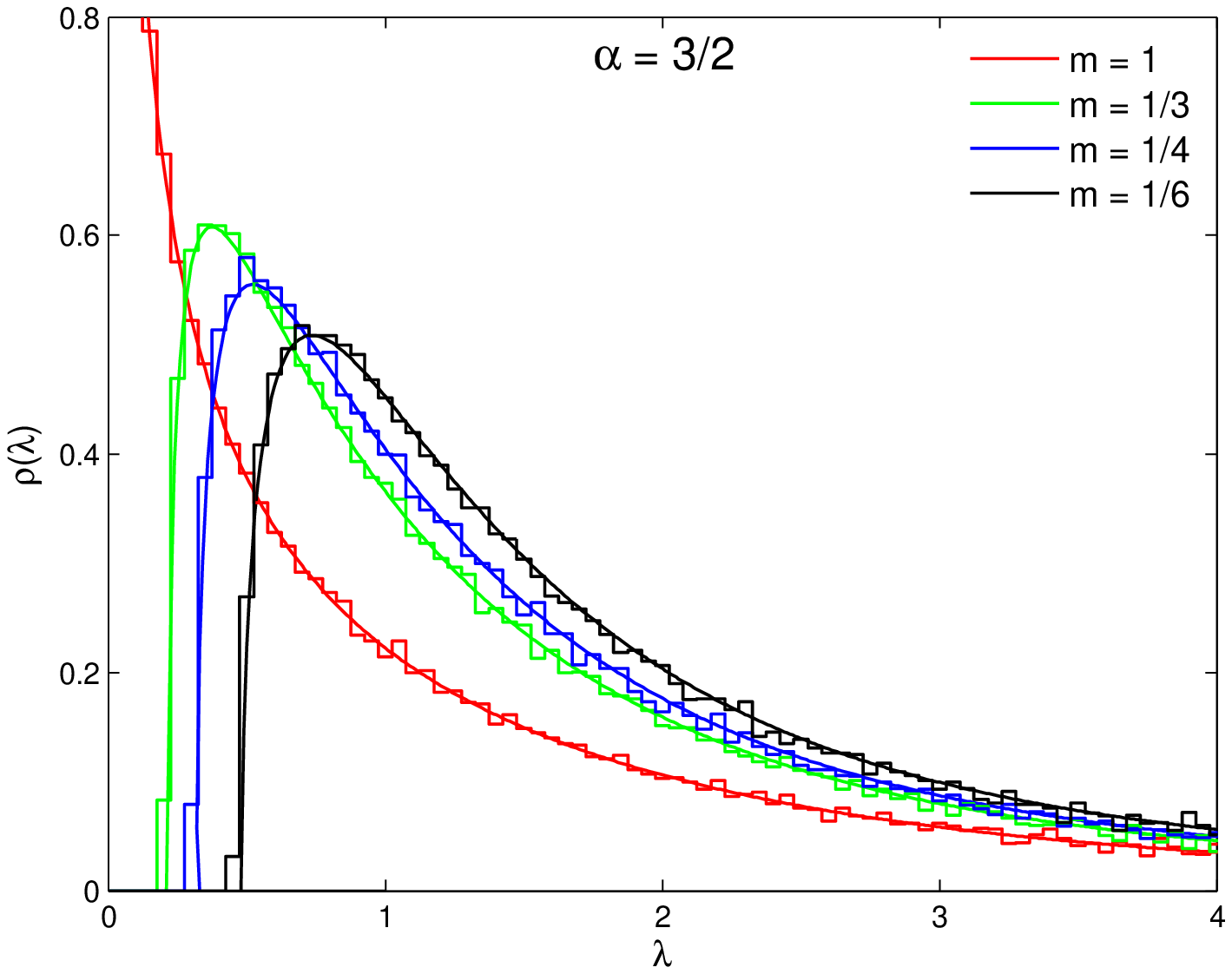}
\includegraphics[width=0.49\textwidth]{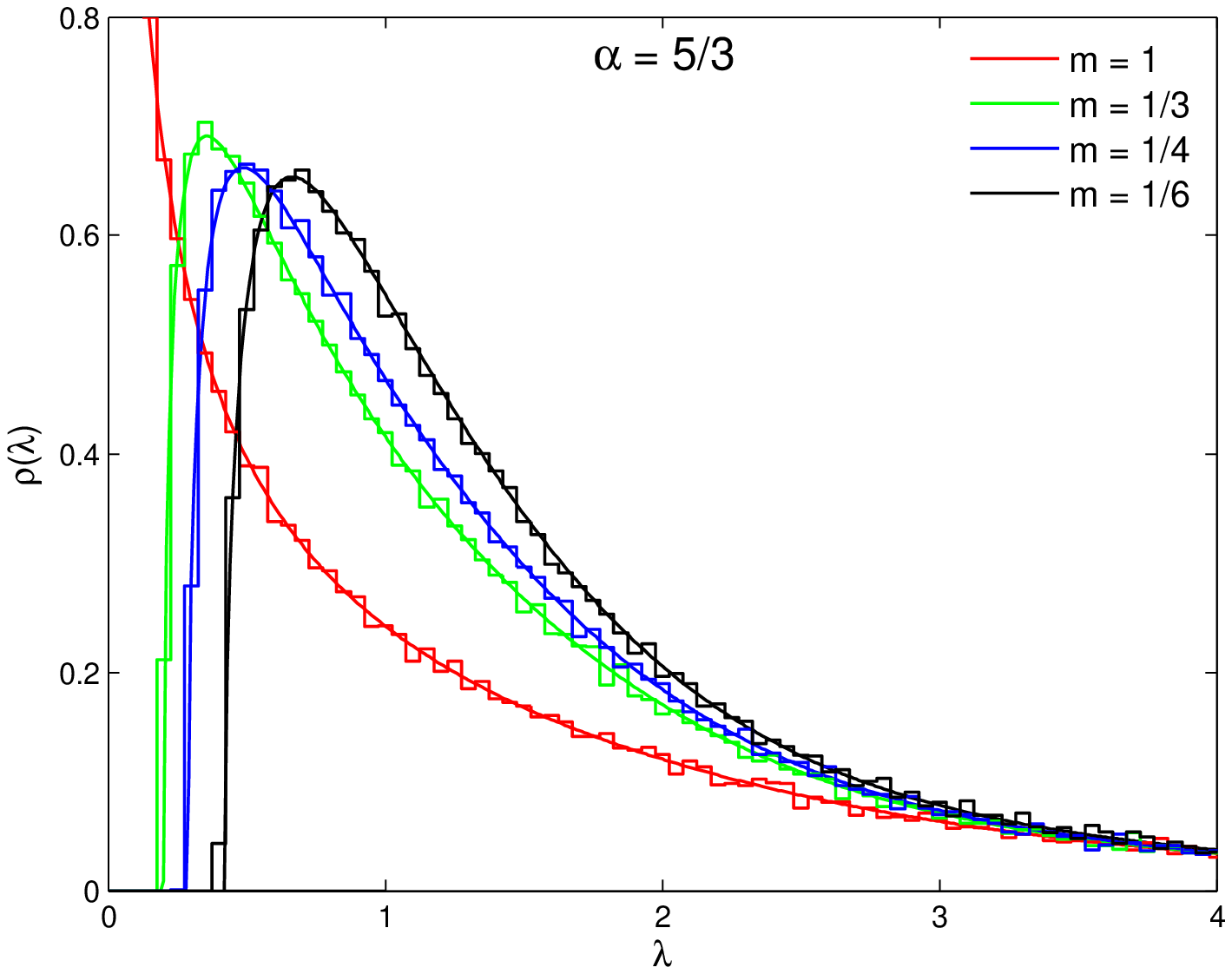}
\end{center}
\caption{\label{fig:MCvalidation}
Spectral densities from the numerical solution of the analytic equation (solid
lines) and from Monte Carlo simulation (stairs). In each case the dimension of
$\mathbf{C}$ is $N = 200$, the number of addends in Eq.~(\ref{eq:sum}) is
$R = 20$, and the number of sampled eigenvalues is $S = 36\,000$.}
\end{figure*}

It has already been shown numerically that the theory works in the
Wigner-L\'evy ensemble \cite{Burda2007}. For the Wishart-L\'evy case we
produced free L\'evy stable random matrices $\mathbf{\Lambda}$ of size
$T \times T$ through Eq.~(\ref{eq:sum}); a $N \times N$ principal minor of
$\mathbf{\Lambda\Lambda}^\mathsf{T}$ is a free Wishart-L\'evy matrix
$\mathbf{C}$ with the desired asymmetry ratio $m = N/T \le 1$.
Such a minor results from the action of the projectors $\mathbf{P}$ in
Eq.~(\ref{eq:corrPLLP}). Since a square matrix of size $T$ contains $n =
\lfloor T/N \rfloor$ non-overlapping principal minors of size $N \leq T$, this
procedure can be repeated for the same matrix $\mathbf{\Lambda}$ with different
projectors $\mathbf{P}_i$, where $i = 1,\dots,n$ labels the projector that
selects the rows from $(i-1)N+1$ to $iN$. Especially if $m$ is small, it is
computationally favourable to follow closely Eq.~(\ref{eq:corrPLLP}) by first
building an $N \times T$ matrix $\mathbf{M}_i = \{ \mathbf{P}_i\mathbf{\Lambda}
\}$ made of $N$ rows out of $\mathbf{\Lambda}$, and then forming the product
$\mathbf{C}_i = \mathbf{M}_i \mathbf{M}_i^\mathsf{T}$. The eigenvalues of
$\mathbf{C}_i$ are accumulated in a histogram that gives the spectrum.
This procedure is repeated producing enough matrices $\mathbf{C}_i$ until the
desired statistical accuracy is reached. All plots in
Fig.~\ref{fig:MCvalidation} have been made using an equal number of eigenvalues
for the sake of comparability.
Free stable laws as defined by the Blue function in Eq.~(\ref{eq:stable}) and
the empirical spectra have different normalisations. For the purpose of a
comparison as in Fig.~\ref{fig:MCvalidation}, this is corrected dividing
$\mathbf{M}$ by a factor $\mathrm{\Gamma}(1+\alpha)^{1/\alpha}$, that can be
obtained comparing the asymptotic behaviour of the two spectra.
The Appendix gives the code for the calculation of the spectral density by
Monte Carlo as just described.

This procedure implements the definition of the Wishart covariance matrix based
on a real random rectangular data matrix $\mathbf{M}$. In this paper free
probability theory has been used to provide an analytic equation for the
spectrum of a Wishart matrix with the simplifying assumption that the matrix
$\mathbf{\Lambda}$ on the right hand side of Eq.~(\ref{eq:corrPLLP}) is
symmetric. Therefore $\mathbf{M}$ may contain symmetric elements too, which is
not necessary in the definition of the Wishart ensemble.
However, it is possible to see that this does not affect the properties of
$\mathbf{MM}^\mathsf{T}$. In other words, the symmetrisation introduced for
simplicity in the analytic derivation does not change the original numerical
problem by introducing correlations. Actually, our Monte Carlo scheme does not
symmetrise the matrix $\mathbf{\Lambda}$ obtained from Eq.~(\ref{eq:sum}) and
matches the analytic spectrum.

\section{Summary}

We have explained the justification as well as the mathematical basis with
which free probability theory enters random matrix theory, in particular in the
context of the Wishart matrix ensemble. Since the derivation of the analytic
solution for the spectra of free stable random matrices has not been published
in a self-contained way yet \cite{Burda2001,Burda2002,Burda2004a,Burda2006b,%
Burda2007}, we recollected it in detail. Then we validated numerically with
Monte Carlo calculations the analytic prediction of the eigenvalue spectrum for
free stable Wishart-L\'evy matrices. Overall we find an excellent consistency
between theory and simulation.

\section*{Acknowledgments}
We are thankful to Maciej Nowak, Jerzy Jurkiewicz and Giulia Iori for useful 
explanations. E.S.\ and D.F.\ have been supported by an Italian research grant
PRIN 2006. Visits of E.S.\ in Marburg were funded through a grant by East
Piedmont University. The stay of M.P.\ in Marburg was supported by two DAAD
grants.

\section*{Appendix: Computer codes}

The numerical solution of
Eqs.~(\ref{eq:transcendentalfinal}--\ref{eq:spectrumfinal}) was computed with
\textsc{Mathematica}~6.0 in almost one line:

\medskip
\noindent\textsl{%
$\alpha$ = 3/2;\\
m = 1/3;\\
width = 0.01;\\
$\lambda$max = 5;\\
SOL := 2;\\
$\rho$ = Table$[\{\lambda$,
N$[$Im$[$M/.NSolve$[$-Exp$[$I2$\pi/\alpha]$M$^{2/\alpha}\lambda$ \\
== m$^{2-2/\alpha}$(M+1/m)(M+1), M$]][[$SOL$]]/(\pi\lambda)]\}$,\\
$\{\lambda$, width, $\lambda$max, width$\}]$;\\
ListPlot$[$Abs$[\rho]]$
}

\medskip
The constant \textsl{SOL} is a positive integer that indicates which of the
possible solutions to pick. A value of $\alpha$ not expressed as a fraction of
integers causes a dramatic increase in running time, which otherwise is less
than a minute.

The Monte Carlo approximation of a free stable random matrix $\mathbf{\Lambda}$
described in Sec.~\ref{sec:approximation}, its use to build a free
Wishart-L\'evy matrix $\mathbf{C}$, and the numerical computation of the
eigenvalue spectrum of the latter including the statistical averaging
described in Sec.~\ref{sec:MC} were carried out with \textsc{Matlab}~7.5:

{\scriptsize
\begin{verbatim}
alpha = 3/2; % index of Levy stable distribution
gam   = 1;   % scale parameter of Levy stable distribution
width = .05; % bin width of eigenvalue histogram
N     = 200; % number of time series
T     = 600; % points in each time series; must be >= N.
R     = 20;  % random rotations
S   = 36000; % number of sampled eigenvalues

psi = (T*R*gamma(1+alpha))^(2/alpha); % normalisation factor
rho = []; % set up array of eigenvalues
iS = 0; % initialise normalisation counter

while (iS < S)

    % approximation of a free stable matrix
    L = stabrnd(alpha,0,gam,0,T,T);
    for iR = 2:R 
        [O,U] = qr(randn(T,T)); % O is a random orthogonal matrix
        L = L + O*stabrnd(alpha,0,gam,0,T,T)*O'; 
    end

    % average over covariance matrices
    for i = 1:N:T-N+1
        Mi = L(i:i+N-1,:); % choose N out of T rows from L
        Ci = Mi*Mi'/psi; % normalisation
        rho = [rho eig(Ci)']; % collect the eigenvalues
        iS = iS + N;
        if (iS >= S)
            break;
        end
    end

end

[histrho lrho] = hist(rho,0:width:100); % build the histogram
histrho = histrho/(length(rho)*width) % normalisation
% lrho contains the abscissa and histrho the ordinate
\end{verbatim}
}

On a 2.2 GHz AMD Athlon 64 X2 ``Toledo'' Dual-Core with Fedora Core 7 Linux,
all the Monte Carlo calculations for Fig.~\ref{fig:MCvalidation} together
lasted about 6.6 hours, ranging from less than 2 minutes each for $\alpha = 1,\
m = 1$ to about 47 minutes for $\alpha \neq 1,\ m = 1/6$. The slow step is the
approximation of $\mathbf{\Lambda}$, i.e.\ the first for-loop, while the second
for-loop with the diagonalisation takes from a maximum of 2.5\% of the total
time for $\alpha = 1,\ m = 1$ down to 0.25\% for $\alpha \neq 1,\ m = 1/6$.
This matches the observation, which we made in the range $N =\ $10--800 and for
the values of $\alpha,\ m,\ R,\ S$ reported in Fig.~\ref{fig:MCvalidation},
that the CPU time is approximatively proportional to $T^2 = (N/m)^2$ and lower
for $\alpha = 1$. In this case, corresponding to the Cauchy distribution,
Eq.~(\ref{eq:chambers}) reduces to $X = \gamma\tan\phi$, which requires fewer
operations than the general formula.

\bibliography{paper}

\begin{thebibliography}{52}

\bibitem{Wishart1928}
J.~Wishart, Biometrika \textbf{20A}, 32 (1928)

\bibitem{Wigner1955}
E.P. Wigner, Ann. Math. \textbf{62}, 548 (1955)

\bibitem{Wigner1958}
E.P. Wigner, Ann. Math. \textbf{67}, 325 (1958)

\bibitem{Guhr1998}
T.~Guhr, A.~M\"uller-Groeling, H.A. Weidenm\"uller, Phys. Rep. \textbf{299},
  189 (1998)

\bibitem{Forrester2003}
P.~Forrester, N.~Snaith, V.~Verbaarschot, J. Phys. A-Math. Gen. \textbf{36}, R1
  (2003)

\bibitem{Foschini1998}
G.J. Foschini, M.J. Gans, Wireless Pers. Commun. \textbf{6}, 311 (1998)

\bibitem{Moustakas2005}
A.L. Moustakas, S.H. Simon, A.M. Sengupta, Acta Phys. Pol. B \textbf{36}, 2719
  (2005)

\bibitem{Bachmat2006}
E.~Bachmat, D.~Berend, L.~Sapir, S.~Skiena, N.~Stolyarov, J. Phys. A-Math. Gen.
  \textbf{39}, L453 (2006)

\bibitem{Steffen2008}
J.~Steffen, J. Air Transp. Manag. \textbf{14}, 146 (2008)

\bibitem{Mehta2004}
M.~Mehta, \emph{Random Matrices}, 3rd~edn. (Elsevier, Amsterdam, 2004)

\bibitem{Edelman2005}
A.~Edelman, Acta Numer. \textbf{14}, 233 (2005)

\bibitem{Laloux1999}
L.~Laloux, P.~Cizeau, J.P. Bouchaud, M.~Potters, Phys. Rev. Lett. \textbf{83},
  1467 (1999)

\bibitem{Plerou1999}
V.~Plerou, P.~Gopikrishnan, B.~Rosenow, L.A.N. Amaral, H.E. Stanley, Phys. Rev.
  Lett. \textbf{83}, 1471 (1999)

\bibitem{Bouchaud2000}
J.P. Bouchaud, M.~Potters, \emph{Theory of Financial Risk and Derivative
  Pricing} (Cambridge University Press, Cambridge, 2000)

\bibitem{Bouchaud2004}
J.P. Bouchaud, Y.~Gefen, M.~Potters, M.~Wyart, Quant. Finance \textbf{4}, 176
  (2004)

\bibitem{Tola2008}
V.~Tola, F.~Lillo, M.~Gallegati, R.N. Mantegna, J. Econ. Dyn. Control
  \textbf{32}, 235 (2008)

\bibitem{Daly2008}
J.~Daly, M.~Crane, H.J. Crane, Physica A \textbf{387}, 4248 (2008)

\bibitem{Markowitz1952}
H.M. Markowitz, J. Finance \textbf{7}, 77 (1952)

\bibitem{Sharpe1964}
W.F. Sharpe, J. Finance \textbf{19}, 425 (1964)

\bibitem{Ormerod2008}
P.~Ormerod, Economics E-Journal \textbf{2}, 26 (2008)

\bibitem{Marchenko1967}
V.A. Mar\v{c}enko, L.A. Pastur, Math. USSR-Sb. \textbf{1}, 457 (1967)

\bibitem{Edelman1988}
A.~Edelman, SIAM J. Matrix Anal. Appl. \textbf{9}, 543 (1988)

\bibitem{Bai1999}
Z.D. Bai, Statist. Sci. \textbf{9}, 611 (1999)

\bibitem{Clauset2007}
A.~Clauset, C.R. Shalizi, M.E.J. Newman, \emph{Power-law distributions in
  empirical data} (2007), \texttt{arXiv:0706.1062}

\bibitem{Cizeau1994}
P.~Cizeau, J.P. Bouchaud, Phys. Rev. E \textbf{50}, 1810 (1994)

\bibitem{Burda2001}
Z.~Burda, J.~Jurkiewicz, M.A. Nowak, G.~Papp, I.~Zahed, Physica A \textbf{299},
  181 (2001)

\bibitem{Burda2002}
Z.~Burda, R.A. Janik, J.~Jurkiewicz, M.A. Nowak, G.~Papp, I.~Zahed, Phys. Rev.
  E \textbf{65}, 021106 (2002)

\bibitem{Burda2003}
Z.~Burda, J.~Jurkiewicz, M.A. Nowak, G.~Papp, I.~Zahed, Acta Phys. Pol. B
  \textbf{34}, 4747 (2003)

\bibitem{Burda2004a}
Z.~Burda, A.~Goerlich, A.~Jarosz, J.~Jurkiewicz, Physica A \textbf{343}, 295
  (2004)

\bibitem{Burda2004b}
Z.~Burda, J.~Jurkiewicz, M.~Nowak, G.~Papp, I.~Zahed, Physica A \textbf{343},
  694 (2004)

\bibitem{Burda2004c}
Z.~Burda, J.~Jurkiewicz, Physica A \textbf{344}, 67 (2004)

\bibitem{Burda2005}
Z.~Burda, J.~Jurkiewicz, B.~Waclaw, Phys. Rev. E \textbf{71}, 026111 (2005)

\bibitem{Burda2006a}
Z.~Burda, A.~Goerlich, B.~Waclaw, Phys. Rev. E \textbf{74}, 041129 (2006)

\bibitem{Burda2006b}
Z.~Burda, A.~Jarosz, J.~Jurkiewicz, M.A. Nowak, G.~Papp, I.~Zahed,
  \emph{Applying free random variables to random matrix analysis of financial
  data} (2006), \texttt{arXiv:physics/0603024}

\bibitem{Burda2007}
Z.~Burda, J.~Jurkiewicz, M.A. Nowak, G.~Papp, I.~Zahed, Phys. Rev. E
  \textbf{75}, 051126 (2007)

\bibitem{Bouchaud2007a}
J.P. Bouchaud, L.~Laloux, M.A. Miceli, M.~Potters, Eur. Phys. J. B \textbf{55},
  201 (2007)

\bibitem{Bouchaud2007b}
G.~Biroli, J.P. Bouchaud, M.~Potters, Europhys. Lett. \textbf{78}, 10001 (2007)

\bibitem{Biroli2007}
G.~Biroli, J.P. Bouchaud, M.~Potters, J. Stat. Mech. p. P07019 (2007)

\bibitem{Vivo2007}
P.~Vivo, S.N. Majumdar, O.~Bohigas, J. Phys. A-Math. Theor. \textbf{40}, 4317
  (2007)

\bibitem{Nica2006}
A.~Nica, R.~Speicher, \emph{Lectures on the combinatorics of free probability}
  (Cambridge University Press, Cambridge, 2006)

\bibitem{Mikusinski2005}
L.~Debnath, P.~Mikusinski, \emph{Introduction to Hilbert Spaces with
  Applications}, 3rd~edn. (Academic Press, San Diego, 2005)

\bibitem{Barndorff-Nielsen2002}
O.E. Barndorff-{N}ielsen, S.~Thorbj{\o}rnsen, P. Natl. Acad. Sci. USA
  \textbf{99}, 16568 (2002)

\bibitem{Voiculescu1986}
D.~Voiculescu, J. Funct. Anal. \textbf{66}, 323 (1986)

\bibitem{Bercovici1993}
H.~Bercovici, D.~Voiculescu, Indiana Univ. Math. J. \textbf{42}, 733 (1993)

\bibitem{Janik1997}
R.A. Janik, M.A. Nowak, G.~Papp, I.~Zahed, Acta Phys. Pol. B \textbf{28}, 2949
  (1997)

\bibitem{Bercovici1999}
H.~Bercovici, V.~Pata, Ann. Math. \textbf{149}, 1023 (1999)

\bibitem{Speicher1993}
R.~Speicher, Publ. Res. Inst. Math. Sci. \textbf{29}, 731 (1993)

\bibitem{Pastur2000}
L.A. Pastur, V.~Vasilchuk, Commun. Math. Phys. \textbf{214}, 249 (2000)

\bibitem{Chambers1976}
J.M. Chambers, C.L. Mallows, B.W. Stuck, J. Amer. Statist. Assoc. \textbf{71},
  340 (1976)

\bibitem{McCulloch1996}
J.H. McCulloch, \emph{stabrnd.m: Stable random number generator} (1996),
  \textsc{Matlab} script, \texttt{http://www.econ.ohio-state.edu/jhm/jhm.html}

\bibitem{Diaconis1987}
P.~Diaconis, M.~Shahshahami, Probab. Eng. Inform. Sc. \textbf{1}, 15 (1987)

\bibitem{Voiculescu1991}
D.~Voiculescu, Invent. Math. \textbf{104}, 201 (1991)

\end{thebibliography}
\bibliographystyle{epj}

\end{document}